\documentclass{aa}  

\usepackage{graphicx}
\usepackage{natbib}
\usepackage{txfonts}

\begin{document}

\title{The spin axes orbital alignment of both stars within the
eclipsing binary system V1143\,Cyg using the Rossiter--McLaughlin
effect.\thanks{Based on observations obtained at the Lick Observatory,
which is operated by the University of California.}}

\titlerunning{Spin axes in V1143\,Cyg}

\author{S. Albrecht \inst{1} 
\and S. Reffert \inst{2}  
\and I. Snellen \inst{1} 
\and A. Quirrenbach \inst{2} 
\and D. S. Mitchell \inst{3} 
}

\offprints{S. Albrecht}

\institute{Leiden Observatory, Leiden University,
     P.O.\ Box 9513, NL-2300 RA  Leiden, The Netherlands\\
\email{albrecht@strw.leidenuniv.nl}
\and ZAH - Landessternwarte, K\"{o}nigstuhl 12, D-69117 Heidelberg, Germany
\and California Polytechnic State University, San Luis Obispo, CA 93407, USA
}

\date{Received May 25, 2007; accepted July 30, 2007}
 
\abstract
{
The Rossiter--McLaughlin (RM) effect, a rotational effect in eclipsing
systems, provides unique insight into the relative orientation of
stellar spin axes and orbital axes of eclipsing binary systems.
}  
{
Our aim is to develop a robust method to analyze the RM effect in an
eclipsing system with two nearly equally bright components. This gives
access to the orientation of the stellar rotation axes and may shed
light on questions of binary formation and evolution. For example, a
misalignment between the spin axes and the angular momentum of the
system could bring the observed and theoretical apsidal motion into
better agreement for some systems, including V1143\,Cyg.
}
{
High-resolution spectra have been obtained both out of eclipse and
during the primary and secondary eclipses in the V1143\,Cyg system,
using the 0.6\,m Coud\'e Auxiliary Telescope (CAT) and the
high-resolution Hamilton Echelle Spectrograph at the Lick
Observatory. The Rossiter--McLaughlin effect is analyzed in two ways:
(1) by measuring the shift of the line center of gravity during
different phases of the eclipses and (2) by analysis of the line shape
change of the rotational broadening function during eclipses.
} 
{
We measured the projection of the stellar rotation axes using the
rotation effect for both main-sequence stars in an eclipsing binary
system. The projected axes of both stars are aligned with the orbital
spin within the observational uncertainties, with the angle of the
primary rotation axis $\beta_{p}$~=~$0.3\pm 1.5^{\circ}$, and the
angle of the secondary rotation axis $\beta_{s}$~=~$-1.2\pm
1.6^{\circ}$, thereby showing that the remaining difference between
the theoretical and observed apsidal motion for this system is not due
to a misalignment of the stellar rotation axes. Both methods utilized
in this paper work very well, even at times when the broadening
profiles of the two stars overlap.
} 
{}


\maketitle

\section{Introduction}
\label{sect:introduction}

Eclipsing binaries are great stellar laboratories for gathering
information on stellar surface structure. During eclipses, varying
parts of the stellar disk are obscured, allowing the observer to
gather spatially resolved information. Without eclipses, this
information is difficult to access.
 
The crossing of a companion in front of a rotating star causes a
change in the line profile of the eclipsed star as, for example, it
first covers mainly the part of the stellar surface which is moving
towards the observer. This change in the line profile results in a
change in the center of gravity of the line and therefore in a change
in the measured radial-velocity of the star. The strength and shape of
this rotation effect is a function of the projection of the stellar
axes on the sky, its inclination (for stars with differential
rotation), the projected rotational velocity, the stellar radius, the
radius of the companion, the stellar limb-darkening, and the orbital
parameters of the system.

The rotation effect was first observed by \cite{Rossiter1924} in
$\beta$~Lyrae, and by \cite{McLaughlin1924} in the Algol system. The
theory of the rotation effect is well understood (e.g.
\citeauthor{Kopal1959} \citeyear{Kopal1959}, \citeauthor{Hosokawa1953}
\citeyear{Hosokawa1953}, \citeauthor{Otha2005} \citeyear{Otha2005}, and
\citeauthor{Gimenez2006} \citeyear{Gimenez2006}). In contrast,
observations of the rotation effect in eclipsing binary systems are
rare (e.g. \citeauthor{Hube1982} \citeyear{Hube1982} and
\citeauthor{Worek1996} \citeyear{Worek1996}). Observation and analysis
of the RM effect has recently received renewed interest, caused by the
possibility of observing the spin-orbital alignment for transiting
exoplanet systems (e.g. \citeauthor{Queloz2000} \citeyear{Queloz2000}
and \citeauthor{Winn2006} \citeyear{Winn2006}) and the potential to
observe features of the planetary atmosphere \citep{Snellen2004}.

\begin{figure*}
  \begin{center}
    \includegraphics[width=12cm]{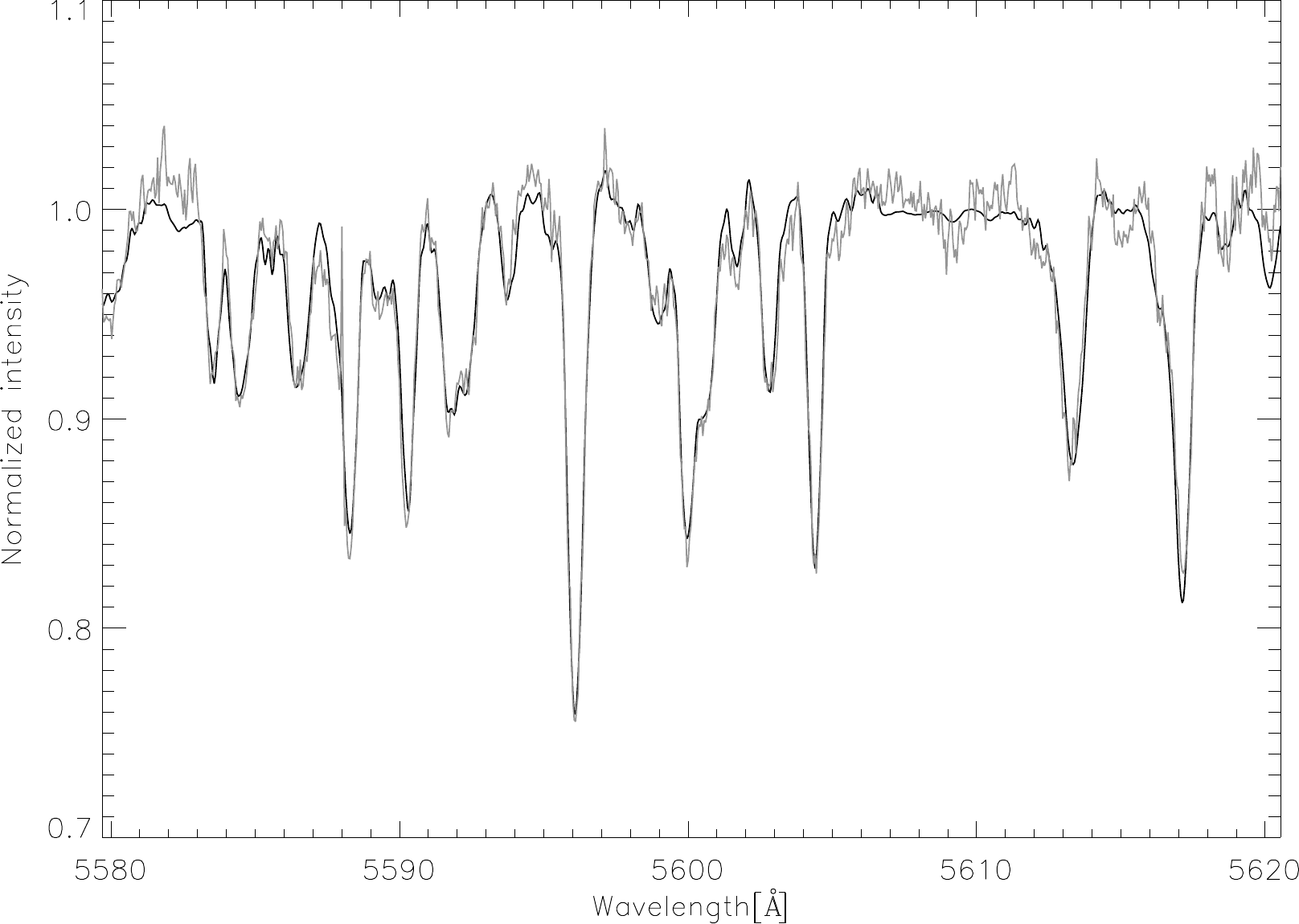}
    \caption {\label{fig:spectrum} Normalized spectrum of V1143\,Cyg
      covering the wavelength range from 5580 to 5620\,\AA. The thin
      line represents the spectrum, and the thick line represents the
      convolution of the narrow-lined template with the BF obtained
      using the SVD algorithm. The quality of the representation of
      the spectrum by the BF and the template is also typical for
      other wavelength regions.}
  \end{center}
\end{figure*}

For the successful observation and interpretation of the rotation
effect in a planetary system, the required S/N and precision in radial
velocity are significantly higher than those required to analyze the
RM effect in a stellar binary system. However, the difficulty in
analyzing a stellar binary system lies in the fact that one has to
deal with the additional light from the eclipsing foreground star. The
spectral lines of the two stars normally blend during the eclipses,
which makes an analysis of the rotation effect in the framework of the
change of the center of gravity during an eclipse
difficult. Nevertheless, the observation of the rotation effect is of
astrophysical interest in binary systems, as it might reveal the
orientation of the stellar rotation axes and provide information about
stellar surface velocity fields. The knowledge of these quantities
might help to answer questions related to binary formation and
evolution, and to the study of apsidal motion.

\begin{table}
  \caption{Parameters for V1143\,Cyg taken from
    \cite{esa1997}$\ddagger$, \cite{Andersen1987}$\dagger$ and
    \cite{Gimenez1985}$\star$. Radius$_{p}$ denotes the radius of the
    primary component and Radius$_{s}$ the radius of the secondary
    component. $L_{s}$/$L_{p}$ denotes the luminosity ratio between
    the secondary and primary.}
  \label{tab:v1143cyg}
  \smallskip
  \begin{center}
    {
      \small
      \begin{tabular}{l l l}
	\hline
	\noalign{\smallskip}
  	HIP & 96620 &  \\ 
	R.A.$_{\rm J2000}$ & $19^{\rm h}38^{\rm m}41^{\rm s}$         & $\ddagger$ \\
	Dec.$_{\rm J2000}$ & $54^{\circ}58^{\prime}26^{\prime\prime}$ & $\ddagger$ \\
	Parallax           & $0\farcs02512(56)$                       & $\ddagger$ \\
	V$_{\mbox{max}}$   & $ 5.89$\,mag                             & $\ddagger$ \\
	Sp. Type           & F\,5\,V                                  & $\dagger$ \\
	Period             & $7\fd6407568(6)$                         & $\star$ \\
	Inclination        & $87.0(1)^{\circ}$                        & $\dagger$ \\
        Radius$_{p}$       & $1.346(23) R_{\odot}$                    & $\dagger$ \\
        Radius$_{s}$       & $1.323(23) R_{\odot}$                    & $\dagger$ \\
	L$_{s}$/L$_{p}$    & $0.96(3)$                                & $\dagger$ \\
	\noalign{\smallskip}
	\hline
      \end{tabular}
    }
  \end{center}
\end{table}

Accordingly, our aim in this research is twofold: I) to develop a
method for deriving information about the orientation of the stellar
rotation axes in an eclipsing system with two nearly equally bright
components; II) to apply it to an astrophysically interesting system,
V1143\,Cyg (e.g.~\citeauthor{Andersen1987} \citeyear{Andersen1987};
\citeauthor{Gimenez1985} \citeyear{Gimenez1985}). V1143\,Cyg (Table
\ref{tab:v1143cyg}) is a bright system consisting of two F5V stars,
and has a high eccentricity ($e=0.54$) that makes it an ideal
candidate for testing a new algorithm. Because of the high
eccentricity, the spectral lines are not as extensively blended during
eclipses. V1143\,Cyg is a young ($2\cdot10^{9}$~yr) system
(\citeauthor{Andersen1987} \citeyear{Andersen1987}), and the measured
apsidal motion, i.e. the precession of the orbit in its own plane
\citep[0.000705$\pm$0.000041$^{\circ}$/cycle,][]{Gimenez1985}, is only
marginally compatible with what is expected theoretically
\citep[0.00089$\pm$0.00015$^{\circ}$/cycle,][]{Andersen1987}. The
precession of the periastron is caused by a general relativistic
effect and a Newtonian contribution, the latter consisting of two
terms which are due to the deformation of the two stars by tides and
stellar rotation. \citeauthor{Andersen1987} (\citeyear{Andersen1987})
suggested the possibility that it could be possible that the tidal
evolution has not yet achieved parallel rotation axes of the stars and
the orbit. This would reduce the expected apsidal motion, thereby
bringing it into a better agreement with the measured apsidal motion.

In the following section we present our observations.
Section~\ref{results} describes the data reduction and the two methods
used to derive the orbital and stellar parameters. The results are
discussed in Section~\ref{sect:discussion}. A summary is given in
Section~\ref{sect:conclusions}.

\section{Observations}
\label{sect:observations}

V1143\,Cyg was observed in the summer/autumn of 2005 and 2006 with the
0.6\,m CAT telescope at the Lick observatory, equipped with the
high-resolution Hamilton Echelle Spectrograph. Observations of the
primary eclipse ($\approx 4$~hours) were made on the night of August
29/30, 2005 with two observations before, nine observations during,
and four observations after the eclipse. The central part of the
secondary eclipse ($\approx 8$~hours) was observed during the nights
August 04/05 and 27/28, 2005. In addition, another 20 observations
were made out of eclipse in order to obtain an accurate orbit
model. During each night, we also observed a set of different radial
velocity standard stars. Before and after every exposure of V1143\,Cyg
($\approx 20$\,min) or a standard star, we made a Thorium-Argon (ThAr)
exposure to obtain a wavelength scale taken close in time to the
observation, minimizing the influence of drifts in the spectrograph on
our measurements. Data from the secondary eclipse taken on August
27/28 had to be discarded from the final analysis, because of an
erroneous ThAr wavelength calibration during that night.

\section{Analysis and results}
\label{results}

For our analysis we used a wavelength range from 4430\,\AA{} to
5750\,\AA, excluding the regions with telluric lines, H$\beta$, and
areas for which an insufficient continuum correction was achieved in
the spectra (usually the edges of the orders). The first steps of the
data reduction consisted of bad-pixel exclusion and continuum
normalization. The latter was carried out by fitting a polynomial of
5$^{\rm th}$ order to the spectra and subsequently dividing the
spectra by it.

In the next step, a weighted mean of the two wavelength scales, based
on the two ThAr spectra obtained directly before and after each
observation, was constructed. The weighting took into account time
differences of the photon weighted midpoint of the observation and the
times of the ThAr exposures. The resulting error in the calculated
radial-velocity due to the drift of the spectrograph during the
observation was estimated to be 75\,m/s on the basis of comparisons of
several ThAr exposure pairs.

\begin{figure} 
  \resizebox{\hsize}{!}{\includegraphics{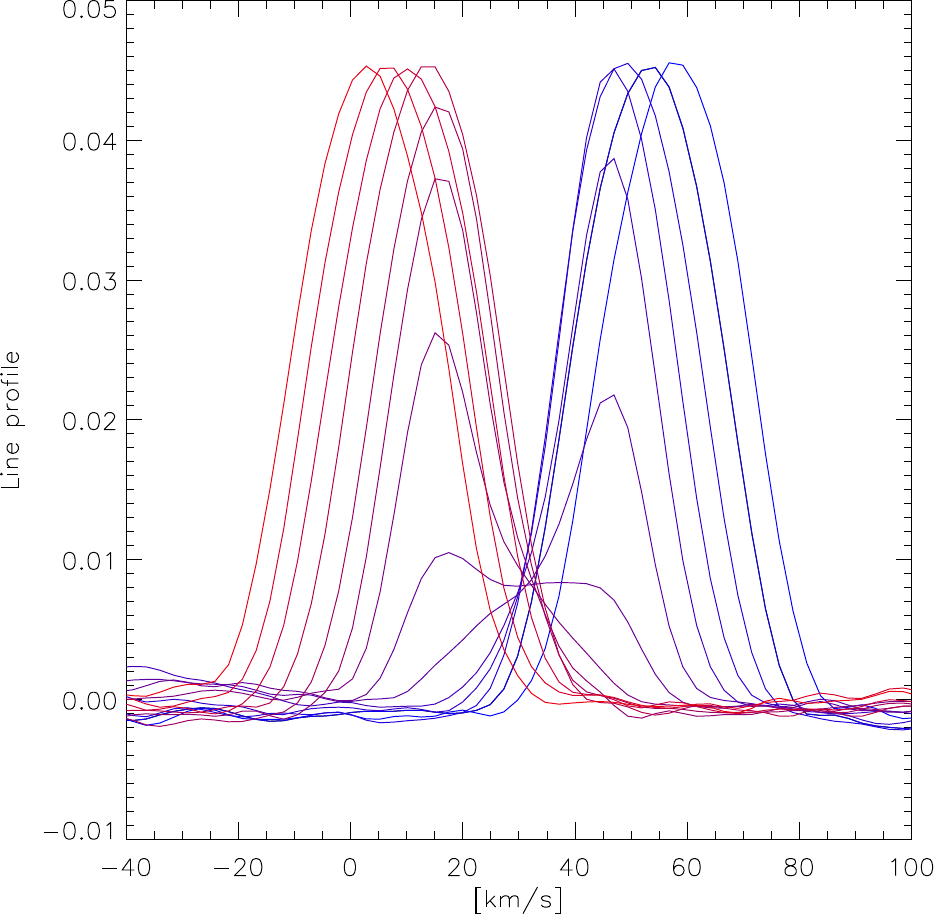}}
  \caption {\label{fig:beta_bf_primary} The panel shows two BFs
    obtained shortly before (higher radial-velocities), nine during,
    and two after the primary eclipse (lower radial-velocities), with
    the secondary BF subtracted, in velocity space. One can see how
    the part of the stellar disk which is moving towards the observer
    is initially obscured. Later in the eclipse a more central part of
    the stellar disk is covered while at the end of the eclipse parts
    of the disk that emit red-shifted light due to the rotation are
    covered. Note that the BF of the primary is moving through
    velocity space. These observations were obtained at $\approx$ 30
    min intervals.}
\end{figure}

The Broadening Function (BF) was calculated to retrieve the line
profiles of the two stars, governed by the velocity fields on the
stellar surface and stellar rotation. The BF represents the function
that projects a narrow-lined template onto the broadened spectrum.  We
calculated the BF using the singular value decomposition (SVD)
technique. This approach is clearly described in \citet{Rucinski1999},
whereas the SVD algorithm itself can be found in
\cite{Press1992}.\footnote{We took all singular values until the first
derivation of the difference between the observed spectrum and the
template broadened by the BF with respect to the singular values used
for the calculation of the BF effectively reached zero.}  In order to
calculate the BF one needs a narrow-lined template; we used the
spectrum of the F7V star HD\,222368 (\citeauthor{Udry1999}
\citeyear{Udry1999}), obtained as one of the radial-velocity standard
stars. Before using the spectrum of HD\,222368 as a narrow-lined
template we deconvolved it using the maximum-likelihood method in
conjunction with a kernel calculated with the same code as applied in
Section~\ref{sect:shape}. An iterative approach was used to identify
the kernel that gave the best results. The most satisfactory results
are achieved when using a kernel with a projected rotational velocity
($v sin i$) of 5\,km/s. Fig.~\ref{fig:spectrum} shows a part of a
typical spectrum obtained from the V1143\,Cyg system. It also shows
for comparison the convolution of the narrow-lined template with the
BF. As V1143\,Cyg is a double-lined binary system with two components
of the same spectral type, the BF itself consists of two peaks that
represent the broadening functions of the two stars, shaped by the
corresponding rotation and velocity fields.

\begin{figure} 
  \resizebox{\hsize}{!}{\includegraphics{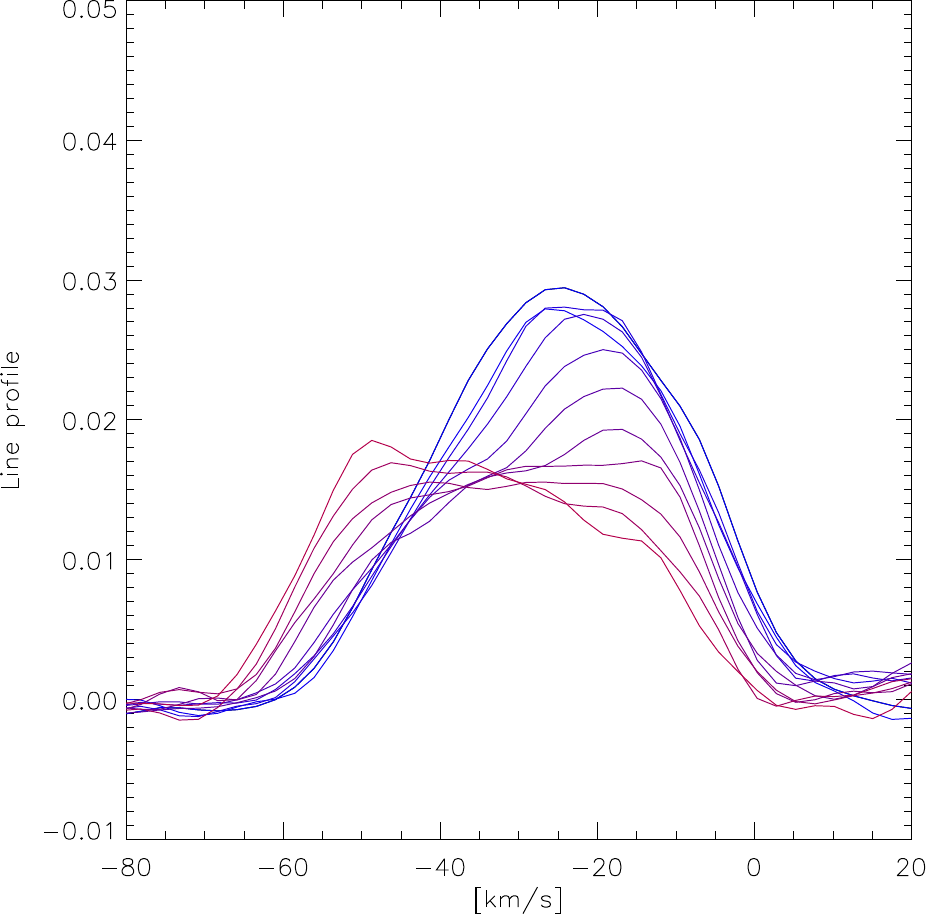}}
  \caption {\label{fig:beta_bf_secondary} The same as
    Fig.~\ref{fig:beta_bf_primary} but for the secondary eclipse. One
    can see BFs of the eleven observations obtained during the
    secondary eclipse with the line profile of the primary
    subtracted. The change in radial-velocity during the secondary
    eclipse is smaller than during the primary eclipse also, the
    coverage of the secondary is less than the coverage of the primary
    during its eclipse. This is due to the orbital inclination not
    being $90^{\circ}$, and the greater distance between the two stars
    along the line of sight during secondary eclipse (See
    Fig.~\ref{fig:orbit_plane}).}
\end{figure}

In this study we are primarily interested in the data taken during the
eclipses and their interpretation in the framework of the RM
effect. Because of the high eccentricity ($e=0.54$) of the system, the
midpoints of the eclipses do not occur when the radial-velocities of
the stars are equal, but shortly before and after the time of equal
velocity, for the primary and secondary eclipses, respectively. During
the middle and the end of the primary eclipse the lines of the two
stars are blended. The same situation occurs during the beginning and
the middle of the secondary eclipse. Hence, although the line blending
is not as severe as in the case of a circular orbit, the light from
the foreground star cannot be ignored.

\begin{figure*}
  \begin{center}
    \begin{tabular}{p{11cm}p{6cm}}
      {     \includegraphics[width=10.9cm]{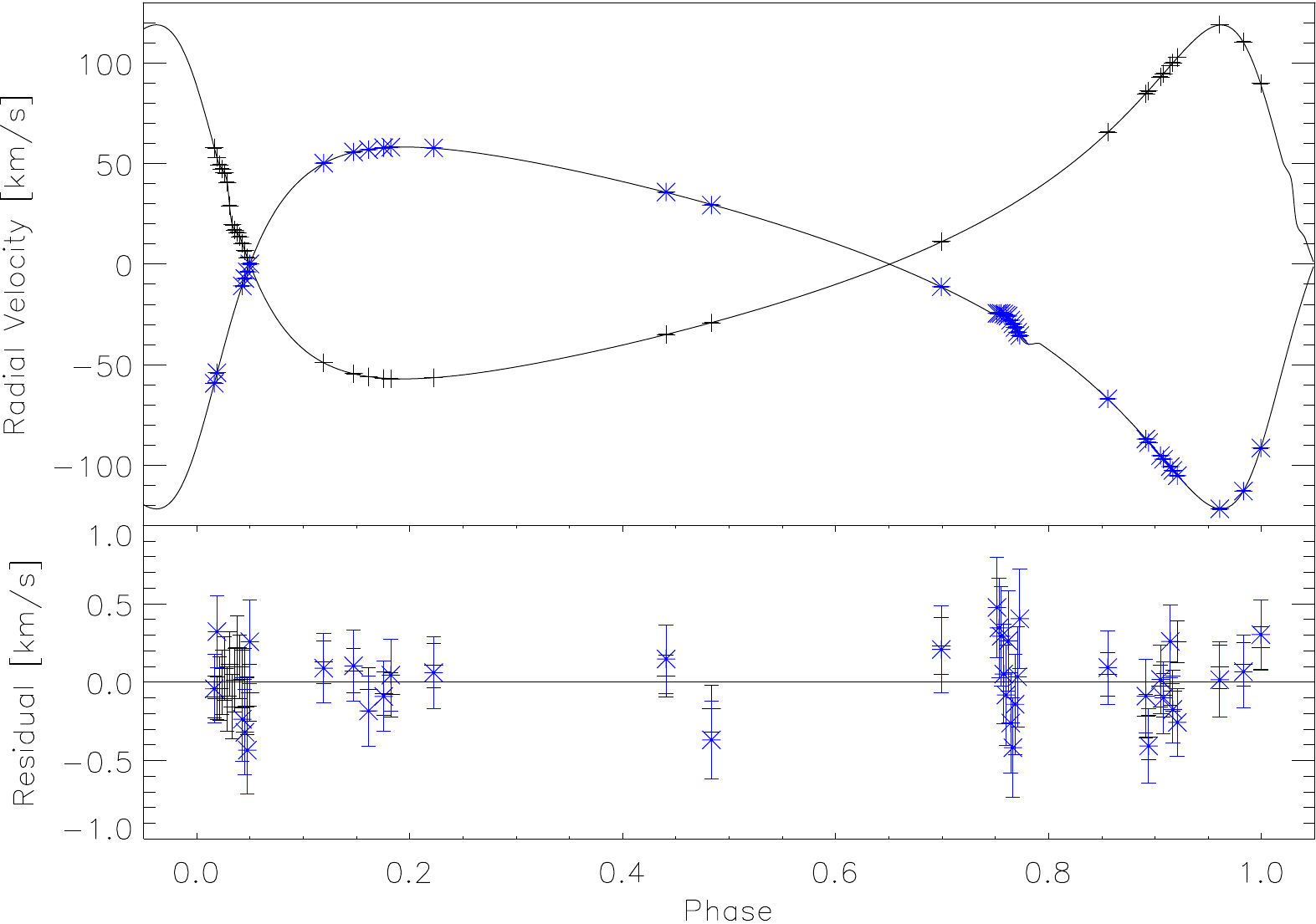}}
      &
      {    \includegraphics[width=5.9cm]{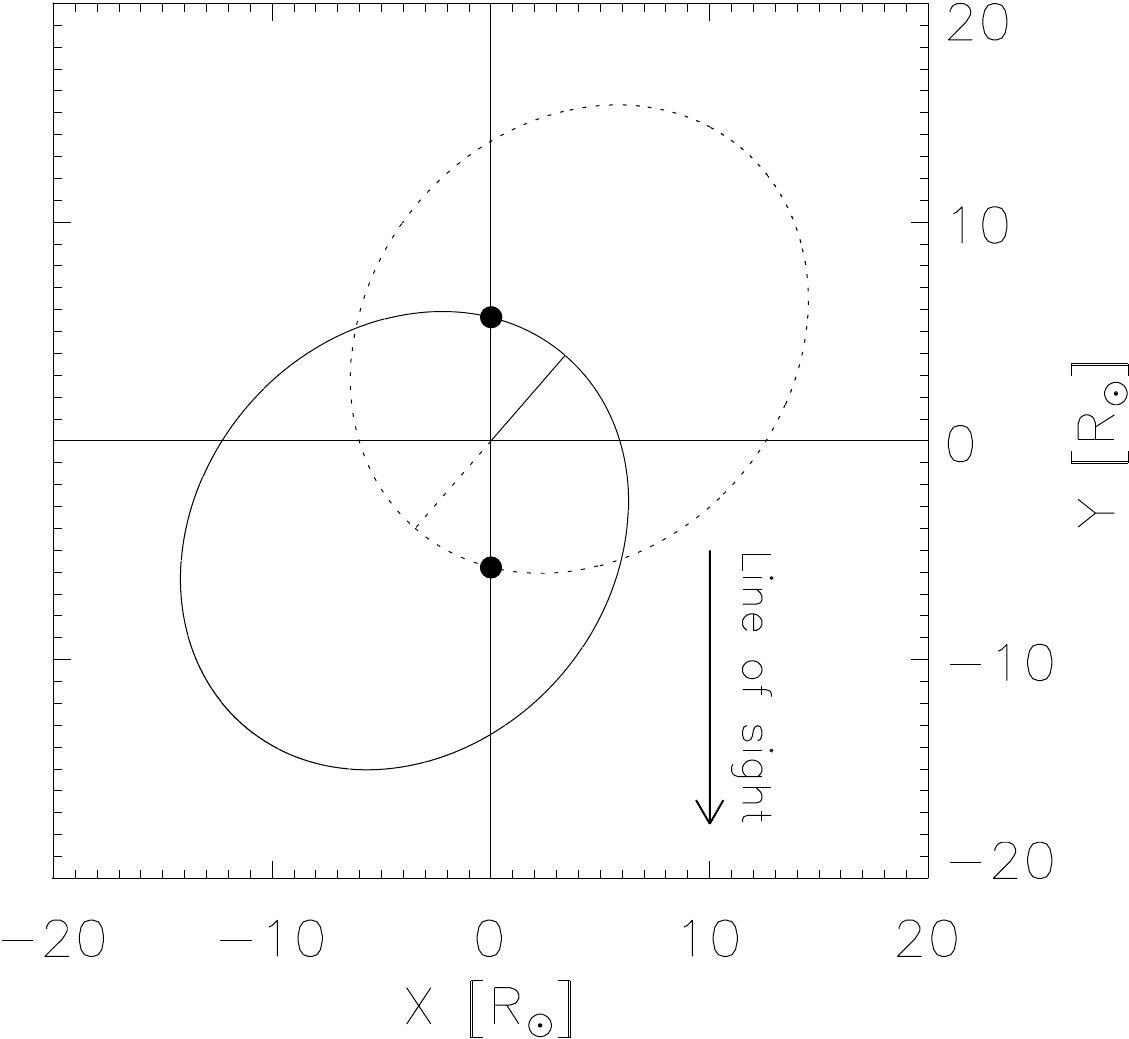}} \\
      \caption{ The radial-velocity measurements of the primary and
	secondary components of V1143\,Cyg, and the orbital solution,
	are plotted against orbital phase. The lower panel shows the
	difference between the best fit and the actual measurements of
	the radial-velocities of the primary and secondary
	component. The midpoint of the primary eclipse occurs at a
	phase of $\approx$ 0.03, and the midpoint of the secondary
	eclipse at a phase of $\approx$ 0.77. Please note that the
	systemic velocity in this graph and in
	Figs.~\ref{fig:beta_primary} and \ref{fig:beta_secondary} is
	already subtracted
	\label{fig:orbit}
      } &
      \caption{The orbit of the binary system V1143\,Cyg shown from
	above the orbital plane. The solid line represents the orbit
	of the primary component and the dashed line the orbit of the
	secondary. The lines from the center of gravity towards the
	orbits indicate the position of the periastron. The big dots
	indicate the positions of the stars at time of mid primary
	eclipse.
	\label{fig:orbit_plane}
      }
    \end{tabular} 
  \end{center}
\end{figure*} 

We followed two different approaches to derive the spin axis of the
eclipsed star with blended lines. The first method is explained in
Section~\ref{sect:center}. We first show how the influence of the
foreground star is subtracted, and subsequently how the center of
gravity from the BF of the eclipsed star is calculated. In
Section~\ref{sect:shape} the BFs of {\it both} stars are used. Here we
do not use the center of gravity of the measured BFs to derive the RM
effect, but the shape of the BFs and their change during the
eclipse. The measured BFs are compared to simulated BFs, and thereby
the parameters that govern the rotation effect are derived.

\subsection{Method 1: The BF's center}
\label{sect:center}

Before determining the parameters involved in the RM effect, an
orbital model has to be obtained. To derive the radial-velocities of
the components out of eclipse, we fitted two Gaussians to the two
peaks in the BFs. A $\chi^{2}$ fit was applied to extract the orbital
parameters. In this work, we adopted the orbital period ($7\fd6407568
\pm 6\times10^{-7}$) for all fits, since it is derived at much higher
accuracy from eclipse photometry than from radial-velocity variations
(\citeauthor{Gimenez1985} \citeyear{Gimenez1985}). The inclination of
the orbit ($87.0 \pm 0.1 ^{\circ}$), and the sizes of the components
$1.346 \pm 0.023 R_{\odot}$ and $1.323 \pm 0.023 R_{\odot}$ have been
adopted from \cite{Andersen1987}; these are needed for later
analysis. The fitted parameters are shown in the second column of
Table~\ref{tab:fit} with their 1-$\sigma$ uncertainties. In addition,
the orbital parameters given by \citet{Andersen1987} and
\cite*{Gimenez1985} are shown for comparison in column five.

The tomography algorithm of \citet*{Bagnuolo1991} is used to
disentangle the primary and secondary spectra. This algorithm uses,
the spectra obtained at different phases of the orbit, and the orbital
parameters of the system, as input. It starts with two synthetic
spectra without spectral lines for the two components in
V1143\,Cyg. For all observations taken outside the eclipses, it shifts
the observed spectra in the rest-frames of each component using the
newly obtained orbital parameters. Subsequently, the synthetic spectra
are compared with the observed spectra. The mean of all the
differences between the synthetic and observed spectra is added to the
synthetic spectra. This complete process is repeated 50 times, but in
our case it converged after only a few iterations.

In the next step, the spectrum of the foreground star is
subtracted. For an observation out of eclipse this is straightforward,
using the spectrum shifted in velocity space to the appropriate
position. During eclipses, one has to incorporate the change in the
light ratio of the two stars due to the eclipses. For this we assumed
a linear limb-darkening law with a limb-darkening coefficient ($u$) of
$0.6$ for both stars. Subsequently, the BF was calculated with only
one star in the spectrum. Figs.~\ref{fig:beta_bf_primary} and
\ref{fig:beta_bf_secondary} show the BFs of the primary and secondary
stars during their eclipses, after subtraction of the foreground
star. Note that, for computational reasons, the continua of the
observed spectra were set to zero and the signs of the spectra have
been changed; this results in positive BFs.

\begin{figure} 
  \resizebox{\hsize}{!}{\includegraphics{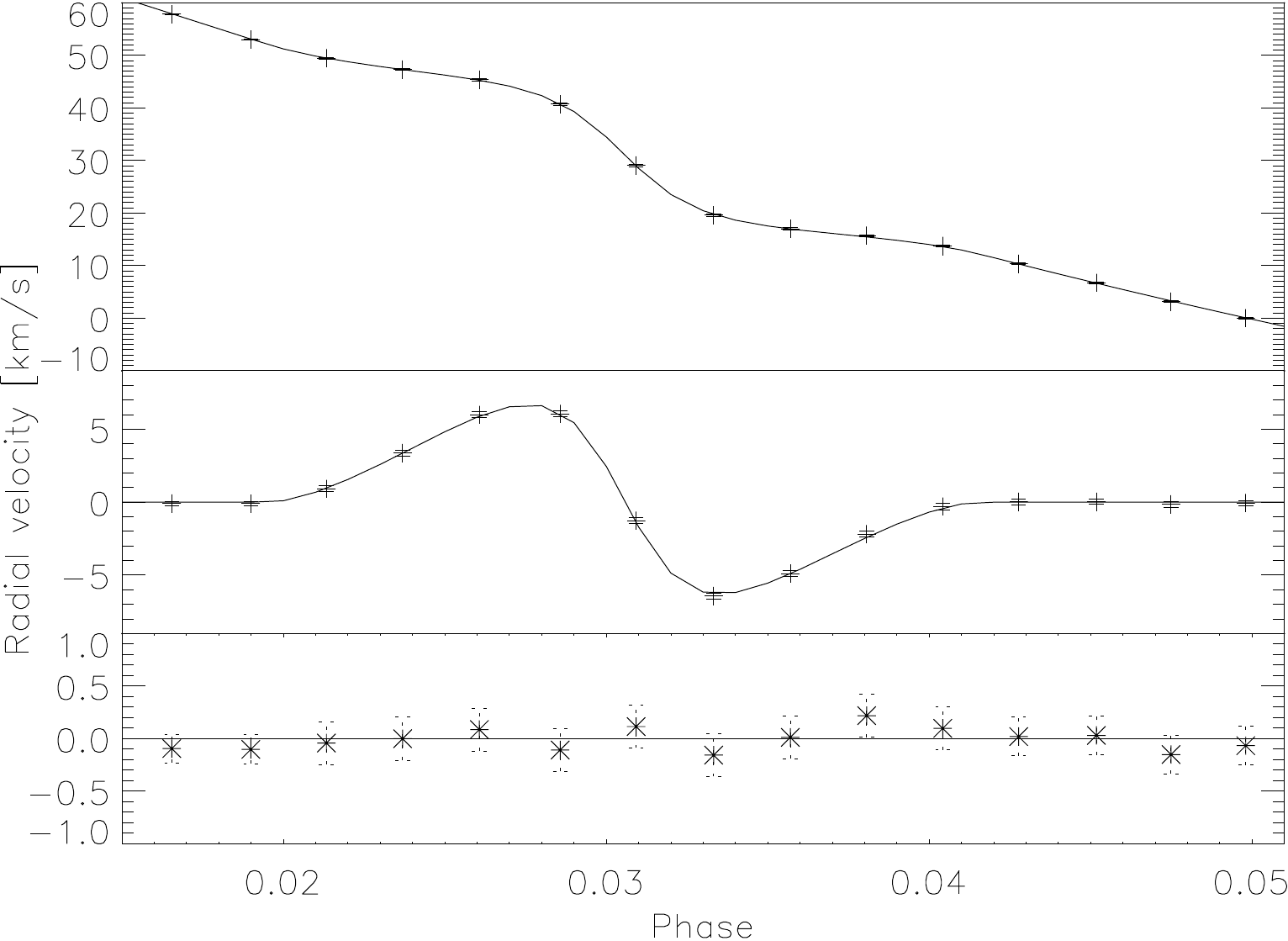}}
  \caption {\label{fig:beta_primary} The rotation effect during the
    primary eclipse. The shift of the center of gravity is plotted
    against the orbital phase. The upper panel shows the measured
    radial velocities along with the best fit. In the second panel the
    radial velocity due to the orbital motion is subtracted from the
    data and the fit. The residuals between data and fit are shown in
    the third panel.}
\end{figure}

\begin{figure} 
  \resizebox{\hsize}{!}{\includegraphics{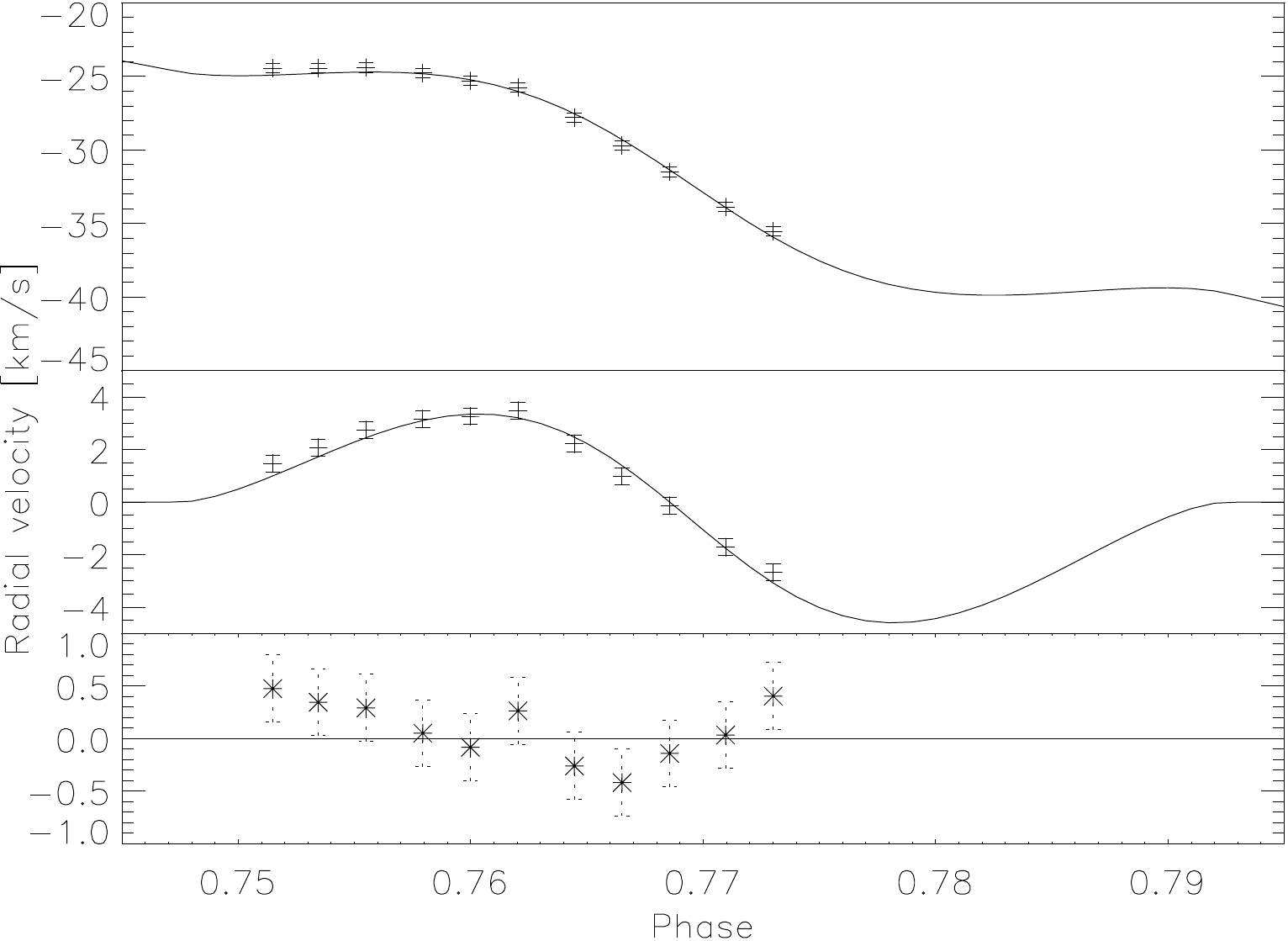}}
  \caption {\label{fig:beta_secondary} The same as for Fig.
    \ref{fig:beta_primary} but this time for the secondary eclipse.}
\end{figure}

The center of gravity of the absorption lines can now be extracted
from the BFs and used to calculate the radial-velocity of the eclipsed
star, including the radial-velocity anomaly introduced by the RM
effect. Using the orbital model and the formula by \cite{Kopal1959}
and \cite{Hosokawa1953}\footnote{Equation 5 has to be adopted for
eccentric orbits.} to calculate the rotation anomaly, the orbital
parameters and the parameters relevant for the RM effect of the two
stars can be derived jointly with a ${\chi}^2$ minimization algorithm;
therefore $v sin i$ and the angles $\beta_{p}$ and $\beta_{s}$ enter
as new parameters. $\beta$ is the angle between the stellar spin axis
projected onto the plane of the sky and the orbital spin axis
projected onto the plane of the sky. $\beta$\,=\,0$^{\circ}$ would
indicate a projected rotation axis perpendicular to the orbital plane,
whereas $\beta$ = 90$^{\circ}$ would indicate that the rotation axis
lies in the orbital plane. The longitude of the ascending node of the
orbit ($\Omega$) is not known, resulting in an ambiguity in the sign
of the angle $\beta$. In our definition, a positive $\beta$ indicates
that the RM effect, integrated over the complete eclipse, would give a
positive residual in radial-velocity. In that case, the companion
spends a longer time in front of the stellar surface moving towards
us, than that part of the stellar surface that is moving away from
us. The linear limb-darkening coefficients of both stars are fixed
during the fits to $0.6$ as they are only weakly constrained by our
fits. In the ${\chi}^2$ fit, all data points out of eclipse and all
data points during the primary and secondary eclipses are fitted
simultaneously. Our radial-velocity data points and the best fit can
be seen in Fig.~\ref{fig:orbit} for the complete orbit including the
two eclipses. Fig.~\ref{fig:orbit_plane} shows the orbit from a point
above the orbital plane. Figs.~\ref{fig:beta_primary} and
\ref{fig:beta_secondary} display the radial-velocities of the primary
and secondary stars during and around their eclipses. The best fit
parameters are listed in the third column of Table~\ref{tab:fit} with
their 1-$\sigma$ uncertainties, as derived from the
${\chi}^2$~fit. All radial-velocity measurements are given with their
epochs in Table~\ref{tab:data} in the Appendix. The radial-velocities
directly after the primary eclipse, and the data point at a phase of
0.7, have been extracted using a Gaussian fit to the BF of one star
after the subtraction of the other star in the spectrum; however,
there is no significant change in the derived parameters if these data
points are omitted. The average uncertainties in the
radial-velocities, out of the eclipses, are 0.15\,km/s for the primary
and 0.25\,km/s for the secondary. The uncertainties of those
radial-velocity measurements that were determined from the centers of
the lines, we estimated to be 0.20\,km/s for the primary and
0.32\,km/s for the secondary. This was calculated by comparing them
with the velocities obtained from a Gaussian fit outside the eclipses.

\subsection{Method 2: Variation of the BF profile}
\label{sect:shape}

In our second approach, to derive information about the binary orbit
and the orientation of the rotation axes, we simulated the shape of
the BF of the two stars, as governed by the orbital motion, stellar
rotation, orientation of the stellar spin axes, velocity fields on the
stellar surface, limb-darkening and, in case of a measurement taken
during an eclipse, the fraction of the stellar disk which is covered
by the companion at the time of the measurement. These simulated BFs
were subsequently compared to the measured BFs, and the relevant
parameters determined. With this method, we not only use the first
moment of the stellar absorption lines during the analysis, but we
utilize the complete BF and its change over the course of the eclipse
as a diagnostic tool for determining the orbital and stellar
parameters.

We simulated the rotation profile of the two stars in the V1143\,Cyg
system with a few thousand elements across the visible half-spheres
with equal surface brightness. We included linear limb-darkening,
asymmetric macro-turbulence and solar-like differential rotation (See
\citeauthor{Gray2005} \citeyear{Gray2005}). Accordingly, besides the
parameters which were fitted in Section~\ref{sect:center}, another
parameter, the Gaussian width of the macro-turbulence, is required. We
kept the $\sigma$ of the Gaussian for the tangential and
radial-velocity fields $\zeta_{RT}$ and their covered surface fraction
equal, as the quality of the fit did not improve by including them as
free parameters. The inclination of the orbit and the sizes of the two
stars can be varied in our fits. However, in the final fits presented
here, we used a linear limb-darkening coefficient of 0.6, solid body
rotation, and the inclination and radii from the literature (see also
Section~\ref{sect:discussion}).

\begin{figure*} 
  \begin{center}
    \includegraphics{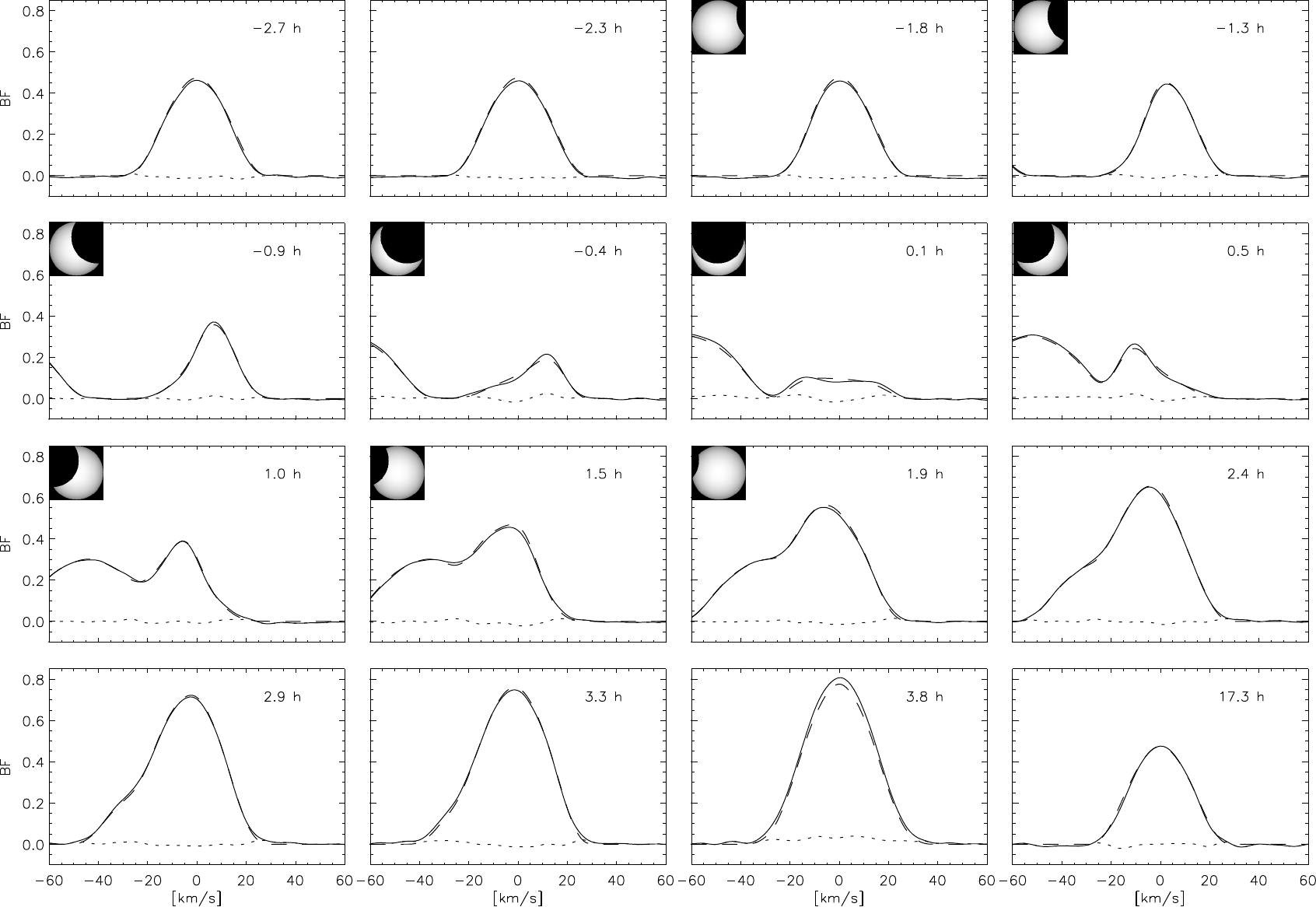}
    \caption {\label{fig:bf_primary}BFs during the primary eclipse
      centered in the frame of the primary. Each panel represents one
      measurement. The solid line represents the measured BF, the
      dashed line shows the best fit BF, while the dotted line
      represents the difference between the fit and the
      measurement. The numbers in the upper right corner of each
      observation indicate the time difference between the photon midpoint of
      the observation and the midpoint of the eclipse in hours. In the
      upper left corner one can see the uncovered part of the primary
      half-sphere for the photon midpoint of that measurement if the star was
      eclipsed during this measurement. One can see that the BF of the
      secondary moves closer to the primary BF in velocity space over
      the course of the eclipse. Please note that the scale of the
      ordinate is arbitrary. The panel in the lower right corner shows
      the BF of the primary for an observation out of eclipse.}
  \end{center}
\end{figure*}

Each exposure of V1143\,Cyg had a duration of
$\approx$\,20\,min. Provided that one takes the photon midpoint of the
observation, this poses no serious problem for the analysis of the
center of gravity of the lines. However, looking at the line profile,
one has to take two effects into account. First, the radial-velocities
of the two stars change during a 20\,min exposure. This effect is
strongest in our data set at the time of the primary eclipse, where
the change in radial-velocity is $\approx$ 3\,km$/$s during one
exposure. This artificially widens a BF taken during a 20\,min
exposure relative to a BF which would have been taken
instantaneously. During the eclipses, a second effect becomes
important; the coverage of the eclipsed stars can change considerably
during 20-min. Again, this effect is stronger during the primary
eclipse than during the secondary eclipse, as the primary eclipse is
deeper and shorter. Sine we need an exposure over several minutes to
receive a sufficiently high S/N ratio\footnote{Out of eclipse the
spectra have a S/N between 40 and 50.}, we introduced the same
``smearing'' in the simulated BFs. We calculated three BFs for each
exposure and stacked them together: one simulated observation 6.6
minutes before the photon midpoint, one at the photon midpoint and one
6.6 minutes after the photon midpoint. We convolved each simulated BF
with a Gaussian of $\sigma$~=~2.5~km/s, representing roughly the
resolution of the spectrograph in the wavelength range used.

\begin{figure*}
  \begin{center}
    \includegraphics{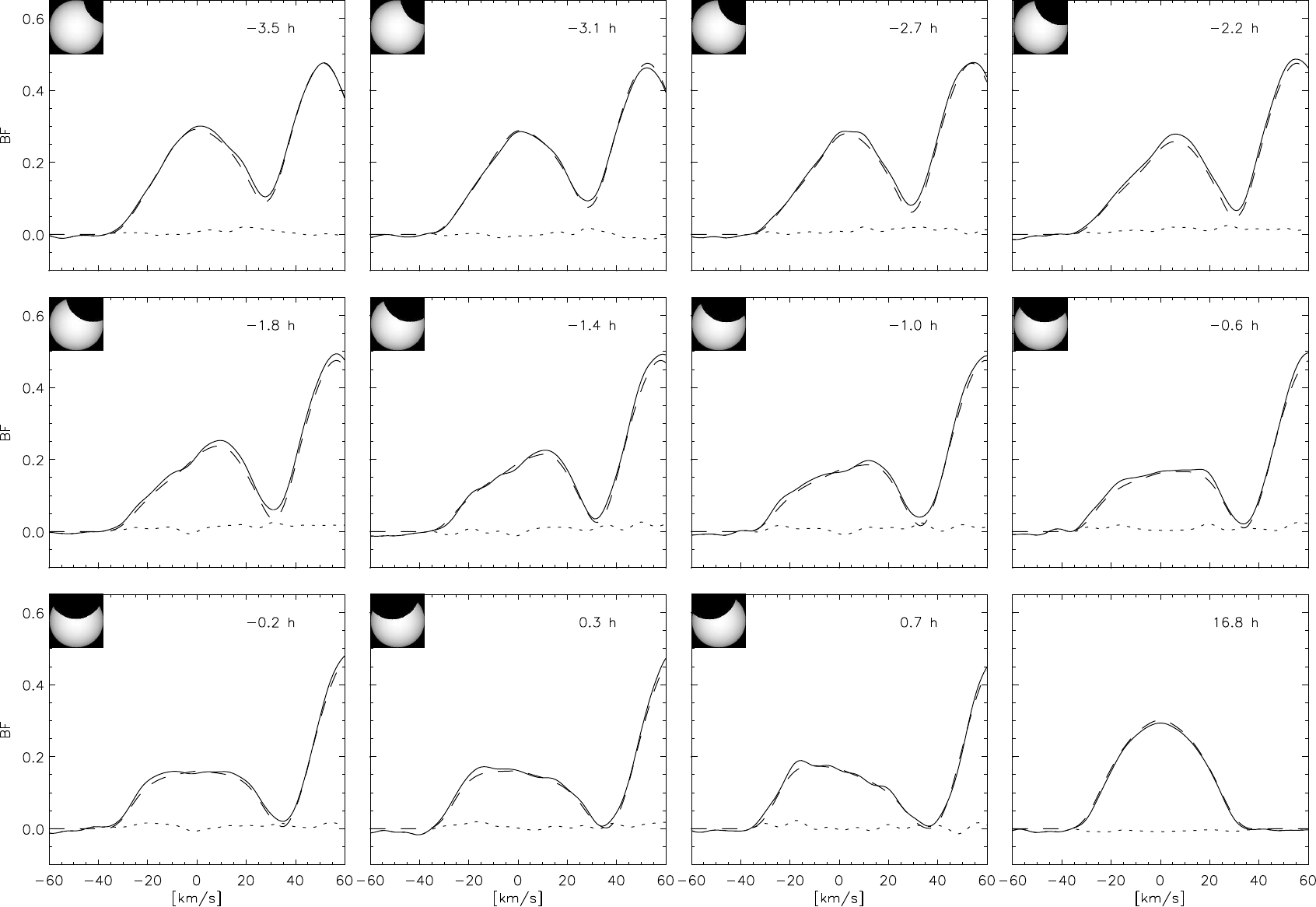}
    \caption {\label{fig:bf_secondary} The same as
      Fig.~\ref{fig:bf_primary} but for the secondary eclipse. One can
      see that the BF of the primary moves further away from the
      secondary BF in velocity space over the course of the
      eclipse. The panel in the lower right corner shows the BF of the
      secondary for an observation out of eclipse. }
  \end{center}
\end{figure*}

In addition to the parameters describing the binary system, one
parameter is included that scales all simulated BFs to the observed
BFs. The primary star has a slightly higher luminosity than the
secondary. Also, the template used might fit one of the two stars
better than the other one. Therefore, we included another parameter
scaling the height of the kernels of the two stars relative to each
other. As mentioned in the beginning of this section, all spectra have
been normalized. During the eclipses the absolute amount of light
changes. Hence the depths of the absorption lines change relative to
the normalized continuum, and therefore the height of the BFs also
change. This has to be incorporated in the calculations of the BFs. We
performed a fit using all 46 spectra obtained out of eclipse and
during the primary and secondary eclipses. The derived values for the
parameters are given in column four of Table~\ref{tab:fit}. The
measured BFs and the best fits can be seen in
Fig.~\ref{fig:bf_primary} for the primary eclipse and in
Fig.~\ref{fig:bf_secondary} for the secondary eclipse.  The
uncertainties were calculated using the bootstrap method described in
\cite{Press1992}.

\section{Discussion} 
\label{sect:discussion} 

\subsection{Orbital parameters} 
\label{sect:orbital_parameters}

Using the methods described in Section~\ref{sect:center} and
Section~\ref{sect:shape} we obtained values for the orbital parameters
of V1143\,Cyg and for the stellar parameters of the two stars.  The
time of periastron passage of the primary (T), the longitude of the
periastron ($\omega$), and the eccentricity ($e$) were determined in
all three fits, and agree with each other to within the 1-$\sigma$
level. Using the values from \cite{Gimenez1985} for $\omega$
($48.26\pm 0.01^{\circ}$) and the apsidal motion rate
(0.000705$\pm$0.000041$^{\circ}$/cycle), and taking the time
difference into account, one would derive an $\omega$ of
$49.31^{\circ} \pm 0.06^{\circ} $ for the time of our
observations. This is also the value stated in
Table~\ref{tab:fit}. The parameter for the semi-amplitude of the
secondary ($K_{s}$) ($90.0\pm0.1$~km/s) falls outside the 1-$\sigma$
range of the literature value ($91.1\pm0.4$~km/s). This difference can
also be seen in the values for the projected semi-major axis of the
system ($a \sin i$). The uncertainty in the radial-velocity of the
center of gravity of the system ($\gamma$) includes the uncertainty of
the radial-velocity in our template star, HD222368 ($5.6\pm0.3$~km/s,
\citeauthor{Udry1999} \citeyear{Udry1999}). Our calculated masses for
the two components in V1143\,Cyg are
$M_{p}$~=~1.355$\pm$0.004\,$M_{\odot}$ and
$M_{s}$~=~1.327$\pm$0.003\,$M_{\odot}$. These values lie in between
the values calculated by \cite{Andersen1987}
$M_{p}$~=~1.391$\pm$0.016\,$M_{\odot}$ and
$M_{s}$~=~1.347$\pm$0.013\,$M_{\odot}$, and the values given by
\cite{Gimenez1985} $M_{p}$~=~1.33$\pm$0.03\,$M_{\odot}$ and
$M_{s}$~=~1.29$\pm$0.03\,$M_{\odot}$.

\subsection{Stellar parameters} 
\label{sect:stellar_parameters}

The stellar parameters obtained in Section~\ref{sect:center} are
derived by analyzing the shape of the rotation anomaly in the radial
velocity. The method relies on a clean subtraction of the foreground
star from the obtained spectra before calculating the BF's, which
depend on orbital parameters derived from out-of-eclipse measurements,
values for the stellar radii, the stellar limb-darkening and the light
ratio between the two stars during the subtraction process, which have
been taken from the literature.

In Section~\ref{sect:shape}, the change in the shape of the absorption
lines is used instead of the rotation anomaly. 
Looking at Figs.~\ref{fig:bf_primary} and \ref{fig:bf_secondary},
one can see that the simulated BFs are somewhat ``rounder'' during
eclipses than the observed BFs. This can clearly be seen during the
central phase of the primary eclipse (Fig.~\ref{fig:bf_primary} panels
in the second row) and during the secondary eclipse
(Fig.~\ref{fig:bf_secondary}).

\begin{table*}
  \caption{Derived parameters of V1143\,Cyg are given together with
    their formal errors. The reduced $\chi^{2}$ of the orbit fit is
    0.95 and the reduced $\chi^{2}$ of the joint fit is 0.96. The
    average uncertainty in radial-velocity for the primary is
    0.16\,km/s and 0.27\,km/s for the secondary, in the joint fit. For
    comparison, values given by \cite{Andersen1987} and
    \cite{Gimenez1985} are also shown. The second and third columns
    present the parameters obtained in Section~\ref{sect:center}. The
    fourth column shows the parameters derived with the method
    presented in Section~\ref{sect:shape}. For the reasons mentioned
    in Section~\ref{sect:discussion} we consider the values found in
    Section~\ref{sect:shape} to be our best parameters and the formal
    errors given for $v \sin i$ and $\zeta_{RT}$ as too small.}
  \label{tab:fit}
  \smallskip
  \begin{center}
    {
      \small
      \begin{tabular}{c  r@{$\pm$}l r@{$\pm$}l r@{$\pm$}l  r@{$\pm$}l }
	\hline
	\noalign{\smallskip}
	Parameter & \multicolumn{4}{c}{\hspace{4mm}Center} 
        & \multicolumn{2}{c}{\hspace{4mm}Shape} & \multicolumn{2}{c}{\hspace{-6mm}\cite{Andersen1987}$\dagger$}\\
	\noalign{\smallskip}
  	& \multicolumn{2}{c}{\hspace{4mm}Orbit} & \multicolumn{2}{c}{\hspace{4mm}Joint fit}
	& \multicolumn{2}{c}{}                  & \multicolumn{2}{c}{\hspace{-6mm}\cite{Gimenez1985}$\star$}\\
	\noalign{\smallskip}
	\hline
	\noalign{\smallskip}
	T [JD-2400000]         &  53536.130 & 0.002       & 53536.131 & 0.002 &   53536.1317 &  0.0006  & \multicolumn{2}{c}{}\\
	$K_{p}$ [km/s]                  &       88.1 & 0.04        & 88.1    & 0.1        & 88.01   & 0.05   & 88.2    & 0.2$\dagger$\\
	$K_{s}$ [km/s]                  &       90.1 & 0.08        & 90.1    & 0.2        & 89.9    & 0.1    & 91.1    & 0.4$\dagger$\\
	$e$                             &      0.538 & 0.001       & 0.538   & 0.001      & 0.5378  & 0.0003 & 0.540   & 0.003$\dagger$\\
	$\omega$ [$^{\circ}$]           &       49.1 & 0.2         & 49.1    & 0.2        & 49.27   &  0.05  & 49.31   & 0.06$\star$\\   
	$a \sin i$ [R$_{\odot}$]        &      22.67 & 0.03        & 22.67   & 0.03       & 22.64   &  0.02  & 22.78   & 0.08$\dagger$\\
	$\gamma$ [km/s]                 &      $-16.8$ & 0.3       & $-16.8$ & 0.3        & $-16.8$ & 0.3    & $-16.5$ & 0.7$\dagger$\\
	$M_{p} \sin^{3}i$ [$M_{\odot}$] &    1.357 & 0.005 & 1.357 & 0.008   & 1.350      &  0.004  & 1.386  & 0.016$\dagger$\\
	$M_{s} \sin^{3}i$ [$M_{\odot}$] &    1.327 & 0.004 & 1.327 & 0.007   & 1.322      &  0.003  & 1.341  & 0.013$\dagger$\\
	$v \sin i_{p}$ [km/s]           &    \multicolumn{2}{c}{}  & 16.9    & 1.0        & 19.6    & 0.1    & 18      & 2$\dagger$\\
	$v \sin i_{s}$ [km/s]           &    \multicolumn{2}{c}{}  & 28.0    & 5.0        & 28.2    & 0.1    & 27      & 3$\dagger$\\
	$\zeta_{RT}P$[km/s]             &    \multicolumn{2}{c}{}  & \multicolumn{2}{c}{} & 3.4     & 0.1    & \multicolumn{2}{c}{}\\
	$\zeta_{RT}S$[km/s]             &    \multicolumn{2}{c}{}  & \multicolumn{2}{c}{} & 3.3     & 0.1    & \multicolumn{2}{c}{}\\
	$\beta_{p}$[$^{\circ}$]         &    \multicolumn{2}{c}{}  & $0.5$   & 4.0        & $0.3$   & 1.5    & \multicolumn{2}{c}{}\\
	$\beta_{s}$[$^{\circ}$]         &    \multicolumn{2}{c}{}  & $-3.9$  & 4.0        & $-1.2$  & 1.6    & \multicolumn{2}{c}{}\\
	\noalign{\smallskip}
	\hline
      \end{tabular}
    }
  \end{center}
\end{table*}

The agreement between the measured BF and the observed BF can be
improved if $u$ is included in the fit. The derived values for $u$
would be $0.9\pm0.1$ for the primary and $0.8\pm0.1$ for the
secondary. As these values are probably too high
(\citeauthor{Gray2005} \citeyear{Gray2005}) and the derived values for
$\beta$ would not change significantly ($\beta_{p}$~=~$-0.6\pm
1.4^{\circ}$ and $\beta_{s}$~=~$-0.3\pm 1.2^{\circ}$), we kept $u$
fixed to 0.6 in the final fits.

Including solar-like differential rotation, the orbital inclination
and the stellar radii as free parameters in the fits also leads to a
better agreement between data and simulation. However, the derived
differential rotation parameters are negative for both stars; the
angular rotation speed is faster at the poles than at the
equator. Furthermore, the fitted radii are not in agreement with the
literature values; the primary radius is increased relative to the
literature value and the secondary decreased relative to its
literature value. The value for the orbital inclination of V1143\,Cyg
in the literature is reproduced by our fit, but weakly constrained. A
negative differential rotation parameter would give the BFs during the
eclipses a more box-like shape. A bigger difference between the
stellar radii would further improve the agreement between simulation
and data during the central part of the primary eclipse. However, a
bigger difference in stellar radii would also mean a bigger difference
in stellar masses, which would not agree with the nearly equal
semi-amplitudes found in the radial-velocity.

We favor a different explanation for the difference of ``roundness''
between the measured and simulated BFs. Looking at an absorption line
originating from one small area of the stellar disk (i.e. no
rotational broadening), the core of an absorption line is formed
higher up in the stellar atmosphere than the wings (see
\citeauthor{Gray2005} \citeyear{Gray2005}), as the same optical depth
is reached earlier in the core than in the wings of the absorption
line. One can now compare the effect that limb-darkening has on the
core and the wings of the absorption lines. In the core of an
absorption line, a high optical depth is reached in a high layer of
the atmosphere looking down on a stellar atmosphere. This also means
that looking from an angle into the stellar atmosphere (i.e. at the
limb) will result in a small change in the average height at which the
core of the line is formed: Only the outer cooler parts of the
atmosphere are probed. For the wings of the absorption lines the
situation is different; the height at which the wings of the observed
absorption line are formed changes more from the center to the limb of
the stellar disk. Therefore, in going from the center to the limb of a
stellar disk the absorption lines become more box-like. The particular
shape change of an absorption line originating under different angles
relative to the observer depends on the details of the stellar
atmosphere. This would result in a non-linear limb-darkening law. Out
of eclipse, the light is integrated over the whole stellar surface,
but for example, during the central part of the primary eclipse light
is only received from the outer part of the stellar disk (See
Fig.~\ref{fig:bf_primary}, icon in the panel of the third column and
second row). Therefore, the shape change of the absorption lines
originating from different parts of the stellar disk might explain the
difference between the observed BFs and the simulated BFs (e.g.,
\citeauthor{pierce1982} \citeyear{pierce1982},
\citeauthor{balthasar1988} \citeyear{balthasar1988} or
\citeauthor{hadrava2006} \citeyear{hadrava2006}).
 
Using the broadening of about 1000 lines, we obtained macro-turbulence
parameters ($\zeta_{RT}$)~of~3.4~km/s and~3.3~km/s for the primary and
secondary, respectively. For comparison, the macro-turbulence
parameter found for the Sun, as a disk-integrated star, for weak and
moderately strong lines is~$\approx4.0$~km/s \citep{Takeda1995}.

The values obtained for the $v sin i$ of the secondary, using the two
different methods described above, agree with each other and are also
consistent with the literature value. Note that there might be a
systematic difference between the obtained radial-velocities of the
secondary during the secondary eclipse and the fit to the
radial-velocities (Fig.~\ref{fig:beta_secondary}). This might be
caused by too high an uncertainty in the orbital parameters used
during the data analysis in Section~\ref{sect:center}. The values
obtained for the primary $v sin i$ do agree with the literature value
within their 1-$\sigma$ error, but do not agree with each other within
their 1-$\sigma$ errors. The method applied in
Section~\ref{sect:center} uses the amplitude of the RM effect to
derive the value for the projected rotational velocity. It assumes a
linear limb-darkening coefficient. As mentioned in the last paragraph,
this might not be sufficient. Also in Section~\ref{sect:shape} a
linear limb-darkening coefficient is used, but $v sin i$ is derived
not only by the use of the RM effect, but also by the shape of the BF
outside of the eclipse. Therefore, we disregard the value for $v sin
i$ obtained in Section~\ref{sect:center} and take the value of
$19.6$~km/s from Section \ref{sect:shape} as the projected rotational
velocity for the primary star in V1143\,Cyg. Due to the remaining
mismatch between the data and the simulation, which might be due to
the use of too simple a model, we consider the formal errors for
$\zeta_{RT}$ and $v sin i$ as too small.

We would like to point out that the methods used are still in their
infancy. Even with the derived values for the stellar radii, the
orbital inclination, limb-darkening and differential rotation are not
free of errors yet; these parameters are not normally accessible, or
only with great difficulty, via spectroscopic data.

\subsection{Orientation of the rotation axes} 
\label{sect:orientation_of_the_rotation_axes}
 
The main focus of this work is the robust determination of the
orientation of the stellar rotation axes. The values derived for the
projection of the rotation axes onto the plane of the sky are given in
the last two rows of Table~\ref{tab:fit}. The $\beta$s derived from
the two different methods agree to within their errors despite the
different systematic problems of each method. Therefore, the
projections of the rotation axes of both stars are, to within their
uncertainties, perpendicular to a vector that lies in the plane of the
orbit and is perpendicular to the line of sight. What does that mean
for the orientation axes of the stars? It is highly unlikely that the
geometry is such that we see the projection of the rotation axes
perpendicular to the orbital plane and the stellar rotation axes have
a high inclination towards the observer. We therefore conclude that
the stellar rotation axes are normal to the orbital plane, and aligned
with each other and the rotation axis of the system.

This leaves the theoretical prediction of the apsidal motion unaltered
for V1143\,Cyg. Hence, the difference between expected
(0.00089$\pm$0.00015$^{\circ}$/cycle) and measured apsidal motion
(0.000705$\pm$0.000041$^{\circ}$/cycle), which just lies outside the
1-$\sigma$ error bars, is also unchanged. The effect of a misalignment
between the stellar rotation axes and the orbital spin axis on the
apsidal motion has been studied by \cite{kopal1978},
\cite{shakura1985}, \cite{company1988}, and \cite{petrova2003}. The
contribution of the stellar rotation to the advance of the longitude
of the periastron is reduced if the stellar rotation axis is tilted
against the orbit spin axis until finally, when the axis of stellar
rotation lays in the orbital plane, its contribution is half as large
and with the opposite sign as when the stellar and orbital axes would
be parallel. In this situation, it contributes to a retrograde
rotation of the periastron. The contribution of the stellar rotation
to the apsidal motion depends not only on the orientation of the axis,
but also on the square of the angular stellar rotation rate. As one
measures only $v \sin i$, a greater inclination towards the observer
would mean a higher angular stellar rotation rate, and therefore a
greater contribution of the rotation term to the overall apsidal
motion. We calculate that if the rotation axes of both stars would lie
in the orbital plane, but have no inclination towards the observer,
then the complete apsidal motion would be 0.00073$^{\circ}$/cycle. If
the rotation axes would have an inclination towards the observer of
$i$~$\approx 70^{\circ}$ in either of the two stars, or of
$i$~$\approx 60^{\circ}$ in both stars, only then would the expected
and measured apsidal motion be in agreement. The secondary, due to its
higher $v \sin i$, has a larger influence on the rotational term of
the apsidal motion than the primary component. As already pointed out,
it is very unlikely that the stellar and orbital spin axes span a
large angle, while their projections on the sky are, in their
uncertainties, parallel.

Our findings do not support the hypothesis advocated by
\cite{petrova2003}, that a misalignment of the stellar rotation axes
with the orbital spin could bring the theoretical and measured apsidal
motion for a number of binary systems, including V1143\,Cyg, into
better agreement.

Our work has excluded the option of a misalignment between the stellar
rotation axes as a possible explanation for the difference between the
expected and measured apsidal motion in V1143\,Cyg. It is therefore
interesting to look at other possibilities that might explain this
difference. As the apsidal motion constant ($k_{2}$) is an important
source of uncertainty in the calculation of the expected apsidal
motion, a new calculation of the apsidal motion constant for
V1143\,Cyg using modern codes for stellar evolution might be of value.

In our analysis of the orbital data we found no indication of a third
body in the V1143\,Cyg system, whose influence might also alter the
apsidal motion. However, because of the limited coverage in time (one
year) and the limited accuracy in radial-velocity, combined with the
possibility that the orbit of a third body could have a lower
inclination, a third body cannot be excluded.

The alignment of the stellar rotation axes could also set a lower
limit to the age of the system, if the axes were not aligned at the
birth of the system. However, this is not a straightforward argument,
since, during the pre-main-sequence phase, synchronizing forces must
have been larger, due to the larger sizes of the stars. Proper
modeling of the evolution of V1143\,Cyg might reveal whether it had
enough time to align its axes. If the time-span to align the axes is
longer than the lifetime of the system, it would mean that V1143\,Cyg,
with its high eccentricity of $0.54$, was born in this way, with all
spin and orbital axes aligned.

\section{Conclusions} 
\label{sect:conclusions}

We measured the Rossiter--McLaughlin effect for both components in the
binary system V1143\,Cyg. We developed two different methods to derive
the angle $\beta$ between the stellar spin axis projected onto the
plane of the sky and the orbital spin axis projected onto the plane of
the sky.

Using the first method, i.e. by determination of the center of the
broadening function, we showed how it is possible to subtract the
absorption lines of the foreground star during an eclipse from the
spectrum. This made it possible to use the shift of the center of
gravity of the absorption lines as a proxy for the rotation effect,
even in systems with blending occuring during eclipses, and derive
values for $\beta$ and $v sin i$. However, for systems with low
eccentricity, blending of the spectral lines during eclipses is
stronger than in V1143\,Cyg. Therefore, every systematic error in the
tomography or in the subtraction of the foreground spectrum due to the
parameters used in the subtraction, will have a substantial effect on
the value of $\beta$ derived.

The second method presented in this paper, the modeling of the shape
of the broadening functions, avoids the problems of the first method
by taking the influence of the eclipsing star into account
explicitly. This makes the method more suitable for eclipsing binary
systems with low eccentricity (higher blending of the spectral
lines). The use of the complete BFs, instead of the center of gravity
or the Gaussian fit to the line, has the additional advantage that it
makes the fitting of the parameters $\beta$ and $v sin i$ more
precise. However, the derived values for these parameters are not
necessarily free of systematic errors. Further work is needed here to
make it possible to derive information about velocity fields on the
stellar surface represented by macro-turbulence, or differential
rotation. If differential rotation is present it might be possible to
derive some information about the inclination of the rotation axes
towards the observer.

With both methods we determined orbital and stellar parameters which
agree within their uncertainties with values derived in earlier
studies, with the exception of the semi-amplitude of the
secondary. The values derived for the angle $\beta$ agree with each
other. We found that the stellar rotation axes in V1143\,Cyg are
normal to the orbital plane for both components, which leaves the
theoretical predictions for the apsidal motion unchanged. This makes
V1143Cyg\, the first binary system with main-sequence stars for which
the orientations of the rotation axes of both components are
determined.

Even with progress made in the field of apsidal motion in the last
years (e.g. \citeauthor{Claret2002} \citeyear{Claret2002}), it is
interesting to apply the methods discussed here to other binary
systems where there is a stronger disagreement between the observed
and expected apsidal motion (e.g. DI Herculis, \citeauthor{Claret1998}
\citeyear{Claret1998}), in order to investigate whether a
misalignment of the stellar rotation axes can be excluded as a cause
for the disagreement, or a misalignment contributes to the
difference. Furthermore, information about the orientation of the
stellar rotation axes in binary systems could be used to explore the
dependence of the synchronization time scales on the semi-major axis,
eccentricity of the binary system and the stellar type of its
components. This might lead to new insight about the formation and
evolution of binary systems and stellar interiors.

\begin{acknowledgements}

We kindly thank Saskia Hekker for taking part of the observations used
in this study, as well as the staff at the Lick Observatory for their
excellent support during the observations. We would also like to thank
Craig Markwardt for the use of routines taken from his web page at
http://cow.physics.wisc.edu/\verb ~ craigm/idl/idl.html. This research
has made use of the Simbad database located at
http://simbad.u-strasbg.fr/ and the Vienna Atomic Line database (VALD)
located at http://ams.astro.univie.ac.at/vald/.

\end{acknowledgements}


\begin{thebibliography}{29}
\expandafter\ifx\csname natexlab\endcsname\relax\def\natexlab#1{#1}\fi

\bibitem[{{Andersen} {et~al.}(1987){Andersen}, {Nordstrom}, {Garcia}, \&
  {Gim\'{e}nez}}]{Andersen1987}
{Andersen}, J., {Nordstrom}, B., {Garcia}, J.~M., \& {Gim\'{e}nez}, A. 1987,
  \aap, 174, 107

\bibitem[{{Bagnuolo} \& {Gies}(1991)}]{Bagnuolo1991}
{Bagnuolo}, Jr., W.~G. \& {Gies}, D.~R. 1991, \apj, 376, 266

\bibitem[{{Balthasar}(1988)}]{balthasar1988}
{Balthasar}, H. 1988, \aaps, 72, 473

\bibitem[{{Claret}(1998)}]{Claret1998}
{Claret}, A. 1998, \aap, 330, 533

\bibitem[{{Claret} \& {Willems}(2002)}]{Claret2002}
{Claret}, A. \& {Willems}, B. 2002, \aap, 388, 518

\bibitem[{{Company} {et~al.}(1988){Company}, {Portilla}, \&
  {Gimenez}}]{company1988}
{Company}, R., {Portilla}, M., \& {Gimenez}, A. 1988, \apj, 335, 962

\bibitem[{ESA(1997)}]{esa1997}
ESA. 1997, ESA SP-1200

\bibitem[{{Gim{\'e}nez}(2006)}]{Gimenez2006}
{Gim{\'e}nez}, A. 2006, \apj, 650, 408

\bibitem[{{Gim\'{e}nez} \& {Margrave}(1985)}]{Gimenez1985}
{Gim\'{e}nez}, A. \& {Margrave}, T.~E. 1985, \aj, 90, 358

\bibitem[{{Gray}(2005)}]{Gray2005}
{Gray}, D.~F. 2005, {The Observation and Analysis of Stellar Photospheres} (The
  Observation and Analysis of Stellar Photospheres, 3$^{\rm rd}$ Edition, by
  D.F.~Gray.~ ISBN 0521851866.~Cambridge, UK: Cambridge University Press,
  2005.)

\bibitem[{{Hadrava}(2007)}]{hadrava2006}
{Hadrava}, P. 2007, in Astronomical Society of the Pacific Conference Series,
  Vol. 370, Solar and Stellar Physics Through Eclipses, ed. O.~{Demircan},
  S.~O. {Selam}, \& B.~{Albayrak}, 164

\bibitem[{{Hosokawa}(1953)}]{Hosokawa1953}
{Hosokawa}, Y. 1953, \pasj, 5, 88

\bibitem[{{Hube} \& {Couch}(1982)}]{Hube1982}
{Hube}, D.~P. \& {Couch}, J.~S. 1982, \apss, 81, 357

\bibitem[{{Kopal}(1959)}]{Kopal1959}
{Kopal}, Z. 1959, {Close binary systems} (The International Astrophysics
  Series, London: Chapman \& Hall, 1959)

\bibitem[{{Kopal}(1978)}]{kopal1978}
{Kopal}, Z., ed. 1978, Astrophysics and Space Science Library, Vol.~68,
  {Dynamics of Close Binary Systems}

\bibitem[{{McLaughlin}(1924)}]{McLaughlin1924}
{McLaughlin}, D.~B. 1924, \apj, 60, 22

\bibitem[{{Ohta} {et~al.}(2005){Ohta}, {Taruya}, \& {Suto}}]{Otha2005}
{Ohta}, Y., {Taruya}, A., \& {Suto}, Y. 2005, \apj, 622, 1118

\bibitem[{{Petrova} \& {Orlov}(2003)}]{petrova2003}
{Petrova}, A.~V. \& {Orlov}, V.~V. 2003, Astrophysics, 46, 329

\bibitem[{{Pierce} \& {Slaughter}(1982)}]{pierce1982}
{Pierce}, A.~K. \& {Slaughter}, C. 1982, \apjs, 48, 73

\bibitem[{{Press} {et~al.}(1992){Press}, {Teukolsky}, {Vetterling}, \&
  {Flannery}}]{Press1992}
{Press}, W.~H., {Teukolsky}, S.~A., {Vetterling}, W.~T., \& {Flannery}, B.~P.
  1992, {Numerical recipes in FORTRAN. The art of scientific computing}
  (Cambridge: University Press, |c1992, 2nd ed.)

\bibitem[{{Queloz} {et~al.}(2000){Queloz}, {Eggenberger}, {Mayor}, {Perrier},
  {Beuzit}, {Naef}, {Sivan}, \& {Udry}}]{Queloz2000}
{Queloz}, D., {Eggenberger}, A., {Mayor}, M., {et~al.} 2000, \aap, 359, L13

\bibitem[{{Rossiter}(1924)}]{Rossiter1924}
{Rossiter}, R.~A. 1924, \apj, 60, 15

\bibitem[{{Rucinski}(1999)}]{Rucinski1999}
{Rucinski}, S. 1999, in ASP Conf. Ser. 185: IAU Colloq. 170: Precise Stellar
  Radial Velocities, ed. J.~B. {Hearnshaw} \& C.~D. {Scarfe}, 82

\bibitem[{{Shakura}(1985)}]{shakura1985}
{Shakura}, N.~I. 1985, Soviet Astronomy Letters, 11, 224

\bibitem[{{Snellen}(2004)}]{Snellen2004}
{Snellen}, I.~A.~G. 2004, \mnras, 353, L1

\bibitem[{{Takeda}(1995)}]{Takeda1995}
{Takeda}, Y. 1995, \pasj, 47, 337

\bibitem[{{Udry} {et~al.}(1999){Udry}, {Mayor}, {Maurice}, {Andersen},
  {Imbert}, {Lindgren}, {Mermilliod}, {Nordstr{\"o}m}, \&
  {Pr{\'e}vot}}]{Udry1999}
{Udry}, S., {Mayor}, M., {Maurice}, E., {et~al.} 1999, in ASP Conf. Ser. 185:
  IAU Colloq. 170: Precise Stellar Radial Velocities, ed. J.~B. {Hearnshaw} \&
  C.~D. {Scarfe}, 383

\bibitem[{{Winn} {et~al.}(2006){Winn}, {Johnson}, {Marcy}, {Butler}, {Vogt},
  {Henry}, {Roussanova}, {Holman}, {Enya}, {Narita}, {Suto}, \&
  {Turner}}]{Winn2006}
{Winn}, J.~N., {Johnson}, J.~A., {Marcy}, G.~W., {et~al.} 2006, \apjl, 653, L69

\bibitem[{{Worek}(1996)}]{Worek1996}
{Worek}, T.~F. 1996, \pasp, 108, 962

\end{thebibliography}

\appendix 

\section{Data}

\begin{table*}[!ht]
  \caption{Radial velocity measurements of V1143\,Cyg out of eclipse
    and during the primary and secondary eclipse. The photon midpoints
    of the observations are given in the first column, and the radial
    velocities of the primary and secondary are given in the second
    and third column.}
  \label{tab:data}
  \smallskip
  \begin{center}
    {\small
      \begin{tabular}{c  r@{$\pm$}l  r@{$\pm$}l }
	\hline
    \noalign{\smallskip}
    HJD &  \multicolumn{2}{c}{Vel. P}  & \multicolumn{2}{c}{Vel. S}  \\
    \noalign{\smallskip}
    \hline
    \noalign{\smallskip}
     [JD-2400000] &   \multicolumn{2}{c}{[km/s]}  &  \multicolumn{2}{c}{[km/s]}   \\
    \noalign{\smallskip}
   \hline
    \noalign{\smallskip}
       53612.665&$  35.40$&   0.13&$ -81.71$&   0.22\\
       53612.683&$  30.58$&   0.14&$ -76.43$&   0.23\\
       53651.654&$ -71.40$&   0.14&$  27.82$&   0.22\\
       53613.663&$ -76.90$&   0.14&$  33.40$&   0.23\\
       53613.771&$ -78.44$&   0.14&$  34.56$&   0.22\\
       53613.880&$ -79.27$&   0.14&$  35.48$&   0.23\\
       53934.843&$ -79.49$&   0.14&$  35.83$&   0.23\\
       53583.676&$ -78.86$&   0.14&$  35.39$&   0.23\\
       53936.818&$ -57.40$&   0.13&$  13.47$&   0.22\\
       53585.669&$ -51.69$&   0.15&$   6.91$&   0.25\\
       53649.639&$  43.23$&   0.14&$ -89.43$&   0.23\\
       53611.709&$  62.33$&   0.14&$-109.40$&   0.23\\
       53611.728&$  63.65$&   0.14&$-111.21$&   0.24\\
       53611.814&$  70.69$&   0.13&$-117.65$&   0.22\\
       53611.835&$  72.30$&   0.14&$-119.47$&   0.23\\
       53611.884&$  76.29$&   0.14&$-123.12$&   0.23\\
       53611.903&$  77.69$&   0.13&$-125.03$&   0.22\\
       53932.846&$  80.45$&   0.13&$-127.61$&   0.22\\
       53581.675&$  96.71$&   0.14&$-144.12$&   0.24\\
       53650.612&$  88.17$&   0.14&$-135.32$&   0.23\\
       53650.737&$  67.50$&   0.14&$-113.86$&   0.22\\
       53612.865&$ -12.10$&   0.18&$ -33.23$&   0.27\\
       53612.884&$ -15.74$&   0.19&$ -29.58$&   0.28\\
       53612.901&$ -19.29$&   0.18&$ -26.26$&   0.28\\
       53612.919&$ -22.44$&   0.18&$ -22.26$&   0.27\\
       53579.678&$ -11.09$&   0.18&$ -33.60$&   0.28\\
              \multicolumn{5}{c}{}                  \\
       53612.701&$  27.01$&   0.20& \multicolumn{2}{c}{} \\
       53612.719&$  24.88$&   0.20& \multicolumn{2}{c}{} \\
       53612.737&$  22.93$&   0.20& \multicolumn{2}{c}{} \\
       53612.757&$  18.28$&   0.20& \multicolumn{2}{c}{} \\
       53612.775&$   6.63$&   0.20& \multicolumn{2}{c}{} \\
       53612.793&$  -2.84$&   0.20& \multicolumn{2}{c}{} \\
       53612.811&$  -5.44$&   0.20& \multicolumn{2}{c}{} \\
       53612.829&$  -6.75$&   0.20& \multicolumn{2}{c}{} \\
       53612.847&$  -8.71$&   0.20& \multicolumn{2}{c}{} \\
                \multicolumn{5}{c}{}     \\
       53587.717& \multicolumn{2}{c}{} &$ -46.88$&   0.32\\
       53587.732& \multicolumn{2}{c}{} &$ -46.88$&   0.32\\
       53587.748& \multicolumn{2}{c}{} &$ -46.85$&   0.32\\
       53587.767& \multicolumn{2}{c}{} &$ -47.20$&   0.32\\
       53587.782& \multicolumn{2}{c}{} &$ -47.74$&   0.32\\
       53587.798& \multicolumn{2}{c}{} &$ -48.19$&   0.32\\
       53587.816& \multicolumn{2}{c}{} &$ -50.23$&   0.32\\
       53587.832& \multicolumn{2}{c}{} &$ -52.14$&   0.32\\
       53587.848& \multicolumn{2}{c}{} &$ -53.93$&   0.32\\
       53587.866& \multicolumn{2}{c}{} &$ -56.31$&   0.32\\
       53587.882& \multicolumn{2}{c}{} &$ -57.96$&   0.32\\
    \noalign{\smallskip}
    \hline
      \end{tabular}
    }
  \end{center}
\end{table*}

\end{document}